\documentclass[%
 amsmath,
amssymb,
 aps,
twocolumn,
longbibliography,
]{revtex4}

\usepackage{graphicx}
\usepackage{dcolumn}
\usepackage{bm}

\begin{document}

\title{Topological Spaser}
\author{Jhih-Sheng Wu}%
 \email{b91202047@gmail.com}
\author{Vadym Apalkov}
\email{vapalkov@gsu.edu}
\author{Mark I. Stockman}%
 \email{mstockman@gsu.edu}
\affiliation{%
Center for Nano-Optics (CeNO) and Department of Physics and Astronomy, Georgia State University, Atlanta, Georgia 30303
}%

\date{\today}

\begin{abstract}
We theoretically introduce a topological spaser, which consists of a hexagonal array of plasmonic metal nanoshells containing an achiral gain medium in their cores. Such a spaser can generate two mutually time-reversed chiral surface plasmon modes in the $\mathbf K$- and $\mathbf K^\prime$-valleys, which carry the opposite topological charges, $\pm1$, and are described by a two-dimensional $E^{\prime}$ representation of the $D_{3h}$ point symmetry group. Due to the mode competition, this spaser exhibits a bistability: only one of these two modes generates, which is a spontaneous symmetry breaking. Such a spaser can be used for an ultrafast all-optical memory and information processing
   
\end{abstract}

\maketitle





The concept of the surface plasmon amplification by stimulated emission of radiation
(spaser, also called plasmonic nanolaser) \cite{Bergman_Stockman:2003_PRL_spaser, Bergman_Stockman_Spaser_Patent_2009, Stockman_JOPT_2010_Spaser_Nanoamplifier}
has recently been experiencing rapid development. Many different types of spasers have been proposed  \cite{Li_Li_Stockman_Bergman_PRB_71_115409_2005_Nanolens_Spaser, Fedyanin_Opt_Lett_2012_Elecrically_Pumped_Spaser, Berman_et_al_OL_2013_Magneto_Optical_Spaser, Zheludev_et_al_Nat_Phot_2008_Lasing_Spaser} and demonstrated \cite{Noginov_et_al_Nature_2009_Spaser_Observation, Oulton_Sorger_Zentgraf_Ma_Gladden_Dai_Bartal_Zhang_Nature_2009_Nanolaser,  Zhang_et_al_Nature_Materials_2010_Spaser,  Long_et_al_Opt_Expr_2011_Spaser_1.5micron_InGaAs,  Hill_et_al_Opt_Expr_2011_DFB_SPP_Spaser, van_Exter_et_al_PRL_2013_Holy_Array_Spasing, Lu-2014-All-Color_Plasmonic, Xiong_et_al_ncomms5953_2014_Room_Temperature_Ultraviolet_Spaser, Lin_et_al_srep19887_2016_Single_Crystalline_Al_ZnO_UV_Spasers, Gwo_et_al_acsphotonics_7b00184_2017_Low_Threshold_Spasers, Song_et_al_acsphotonics_b01018_2017_Perovskite_Grating_Spaser}. The spasers were also applied to various problems including explosives detection \cite{Zhang_et_al_Nat_Nano_2014_Spaser_Explosives_Detection}, monitoring of the nano-environment \cite{Ma_acsphotonics.7b00438_2017_High_Stability_Spasers_for_Sensing, Ma_et_al_nanoph-2016-0006_Nanophotonics_2017_Spaser_Sensing}, cancer therapeutics and diagnostics (theranostics) \cite{Galanzha_Nat_Comm_Spaser_biological_probe_2017}. 
 
A particular type of the spasers is represented by plasmonic crystals that  include gain media \cite{Zheludev_et_al_Nat_Phot_2008_Lasing_Spaser, Odom_et_al_Nature_Nano_2013_Spasing_in_Strongly_Coupled_Nanoprticle_Array, Stockman_Asia_Mater_2013_am201362a_Lasing_Spaser_in_Photonic_Crystals, van_Exter_et_al_PRL_2013_Holy_Array_Spasing}. Such spasers belong to the class of lasing spasers \cite{Zheludev_et_al_Nat_Phot_2008_Lasing_Spaser}, which are nanostructured plasmonic metasurfaces consisting of a periodic lattice of individual spasers. Due to interactions in the near-field, the individual spasers lock in phase to generate temporaly- and spatially-coherent fields.  However, such a fundamental question as the effects of topological properties (the Berry curvature) \cite{Berry_Phase_Proc_Royal_Soc_1984,  Xiao_Niu_RevModPhys.82_2010_Berry_Phase_in_Electronic_Properties} of the plasmonic Bloch bands of such crystals 
has not yet been investigated. 

Recently, a groundbreaking work has been carried out aimed at  obtaining a topological lasing in a plasmonic-photonic (diffractive) lattice of honeycomb symmetry  \cite{Torma_et_al_PhysRevLett.122.013901_2019_Spasing_in_Honecomb_Lattice}. The plasmonic lasing was observed at the $\mathbf K$-points but only in a mode of the $A_1^\prime$ symmetry, which is a singlet (scalar) representation.  Therefore, this mode does not possess a chiral topological charge. 

\begin{figure}
\includegraphics[width=0.95\columnwidth]{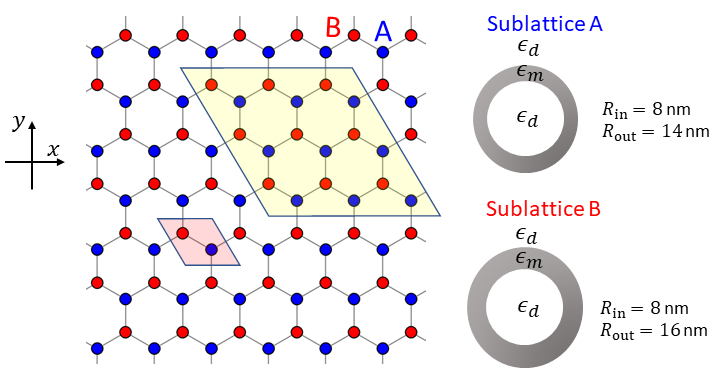}
\caption{Honeycomb array of metal nanoparticles. It consists of two inequivalent sublattices: A and B. Sublattice A consists of plasmonic metal nanoshells with the inner radius of 8 nm and the outer radius of 14 nm while sublattice B consists of similar nanoshells with the inner radius of 8 nm and the outer radius of 16 nm. The gain medium is placed inside the nanoshells. The lattice bond length is set as 50 nm. The primitive unit cell of the honeycomb crystal structure is shown by the red parallelogram. The supercell, which describes the periodicity of the SP's at both the $\mathbf K$- and $\mathbf K^\prime $-points, see Eqs. (\ref{eq:hsp_2s_1})- (\ref{eq:hsp_2s_4}), is marked by the yellow parallelogram.}\label{fig:honeycomb_array}
\end{figure}

In this Letter, we propose a topological spaser that geometrically is a deeply-subwavelength  two-dimensional (2d) crystal (metasurface) with a honeycomb symmetry built of two different triangular sublattices, $\mathrm A$ and $\mathrm B$ -- see Fig.\ \ref{fig:honeycomb_array}.  A metaatom of such a lattice is a plasmonic metal nanoshell containing an achiral gain medium, similar to the spaser geometry of Ref.\ \onlinecite{Stockman_JOPT_2010_Spaser_Nanoamplifier}. The $\mathrm A$- and $\mathrm B$-sublattices of such a topological spaser differ in size and shape of the constituent nanoshells (see the caption to  Fig.\ \ref{fig:honeycomb_array}), so their individual spasers have different eigenfrequencies. Note that a natural example of a honeycomb lattice consisting of two different sublattices is provided by the transition metal dichalcogenide (TMDC) crystals \cite{Heinz_et_al_Nat_Phys_2015_Biexcitons_in_WSe2, Novoselov_et_al_Science_2016_2D_Materials_and_Heterostructures, Basov_et_al_aag1992.full_2016_Polaritons_in_2D}. 

The spasing eigenmodes are surface plasmons (SPs) that should be classified corresponding to irreducible representations of the symmetry point group of the lattice unit cell, which is $D_{3h}$  \cite{Landau_Lifshitz_Quantum_Mechanics:1965}. This group has six representations, of which $E^{\prime}$ is a two-dimensional (doublet) representation with the desired properties -- see Appendix.  It describes two degenerate modes time-reversed to each other, which carry topological charges of $Q_T=\pm1$ defining their chirality; their local fields rotate in time and space in the opposite directions.

For the proposed topological spaser, we will show that the degenerate eigenmodes at the $\mathbf K$- and $\mathbf K^\prime$-points strongly compete with each other, making such a topological spaser bistable: either $Q_T=1$ or $Q_T=-1$ mode can generate.
These spasing modes are topological Berry plasmons \cite{Song_et_al_PNAS_2016_Chiral_plasmons_without_magnetic_field}. However, in Ref.\ \onlinecite{Song_et_al_PNAS_2016_Chiral_plasmons_without_magnetic_field}, the system's chirality was due to  the induced valley polarization. In a sharp contrast, in our case the system is originally achiral and $\mathcal T$-reversible: no integral Berry curvature or magnetic field are present; the chirality is self-organized due to the mode competition 
causing a spontaneous violation of the $\mathcal T$-reversal and $\sigma_v^\prime$-reflection symmetries.


The quasistatic SPs eigenmodes \cite{Bergman_Stroud_1992} are electric potentials,  $\varphi_{\nu \mathbf k} (\mathbf{r})$,   characterized by lattice momentum $\mathbf k$ and band index $\nu$. These  eigenmodes can be found from the quasistatic equation \cite{Stockman:2001_PRL_Localization}
\begin{align}
\nabla\left[\Theta(\mathbf{r})\nabla \varphi_{\nu \mathbf k} (\mathbf{r}) \right]= s_{\nu \mathbf k}\nabla^2\varphi_{\nu \mathbf k} (\mathbf{r}) ,
\label{eq:Bergman}
\end{align}
where $1> s_{\nu \mathbf k}>0$ are the eigenvalues, and $\Theta(\mathbf{r})$ is the characteristic function, which is 1 inside and 0 outside the metal.  These eigenmodes satisfy an orthonormality condition: $\int\nabla\varphi_{\nu \mathbf k} \nabla\varphi_{\nu^{\prime} \mathbf k^{\prime}}^{\ast} d^{3}r =\delta_{\nu \nu^{\prime}}\delta_{\mathbf k \mathbf k^{\prime}}$.

We employ the tight-binding approximation where the on-site states are the modes of isolated metal nanoshells, and the coupling between them is  the nearest-neighbor dipole interaction. For each nanoshell, we consider two eigenmodes constituting a basis for the $E^{\prime}$ doublet representation of the $D_{3h}$ point symmetry group,
$\varphi^{(\alpha)}_m\propto Y_{1m}$, where $Y_{1m}$ are the spherical harmonics, $m=\pm 1$, and  $\alpha = \mathrm A$ or $\mathrm B$. These eigenmodes with $m=\pm1$ have the electric field  components in the $xy$-plane of the lattice and are coupled by the dipole interaction.

The eigenmodes $\varphi^{(\alpha)}_m$ of an isolated nanoshell of a sublattice $\alpha= \mathrm A, \mathrm B$ satisfy the following equation [cf.\ Eq.\ (\ref{eq:Bergman})]
\begin{align}
\nabla\left[\Theta^{(\alpha)} (\mathbf{r})\nabla  \varphi_{m}^{(\alpha)} (\mathbf{r})\right]= s_{m}^{(\alpha)}\nabla^2\varphi_{m}^{(\alpha)} (\mathbf{r})~.
\label{single_nanoshell}
\end{align}
where  $\Theta^{(\alpha)} (\mathbf{r})$ is 1 inside  the metal of nanoshell $\alpha$  and 0 elsewhere, and $\mathbf{r}$ is relative to the center of this nanoshell. The solution of Eq.\ (\ref{single_nanoshell}) can be expressed as an expansion over the spherical harmonics, $Y_{1m}$, see  Appendix. 

With the known eigenmodes of the isolated 
nanoshells, we express the quasistatic potential for a mode with quantum numbers $\nu, \mathbf k$ as a sum over the lattice and $m=\pm1$,
\begin{equation}
\varphi_{\nu \mathbf k}(\mathbf r)=\sum_{j \alpha m}C^{(\alpha)}_{\nu \mathbf k m}\sqrt{s^{(\alpha)}_m} \exp\left(i\mathbf{k}\mathbf R^{(\alpha}_j\right)
\varphi^{(\alpha)}_m\left(\mathbf r-\mathbf R^{(\alpha)}_j\right),
\label{expand}
\end{equation}
where $\mathbf R^{(\alpha)}_j$ is the lattice vector of  the nanoshell $j$ center in  sublattice $\alpha$. Expansion coefficients $C^{(\alpha)}_{\nu \mathbf k m}$ satisfy the following  tight-binding equations 
\begin{equation}
\sum_{\alpha^\prime m^\prime}H_{\alpha m,\alpha^\prime m^\prime}(\mathbf k) C^{(\alpha^\prime)}_{\nu \mathbf k m^\prime}=s_{\nu\mathbf k}C^{(\alpha)}_{\nu \mathbf k m}~.
\label{tight_binding}
\end{equation}
Here the nearest neighbor tight-binding Hamiltonian is
\begin{align}
&H_{\alpha m,\alpha^\prime m^\prime}=\notag  \\
&\begin{cases}
s^{(\alpha)}_m\delta_{mm^{\prime}}, &\alpha =\alpha^{\prime},\\[5pt]
\sqrt{s^{(\alpha)}_m}\sqrt{s^{(\alpha^{\prime})}_{m^{\prime}}}
\displaystyle{\sum_{j^{\prime}}}
\exp[i\mathbf{k}(\mathbf{R}^{\alpha}_{j^{\prime}}-\mathbf{R}^{\alpha}_{j})]\times&\\
\int \nabla\varphi^{(\alpha)*}_m\left(\mathbf r-\mathbf R^{(\alpha)}_j\right) \nabla\varphi^{(\alpha^{\prime})}_m\left(\mathbf r-\mathbf R^{(\alpha^{\prime})}_{j^{\prime}}\right) d^3r
,&\alpha \neq \alpha^{\prime},
\end{cases}
\end{align}
where $j$ is an arbitrary lattice site, and  $j^\prime$ are the nearest-neighbor sites to it.

\begin{figure}
\includegraphics[width=0.95\columnwidth]{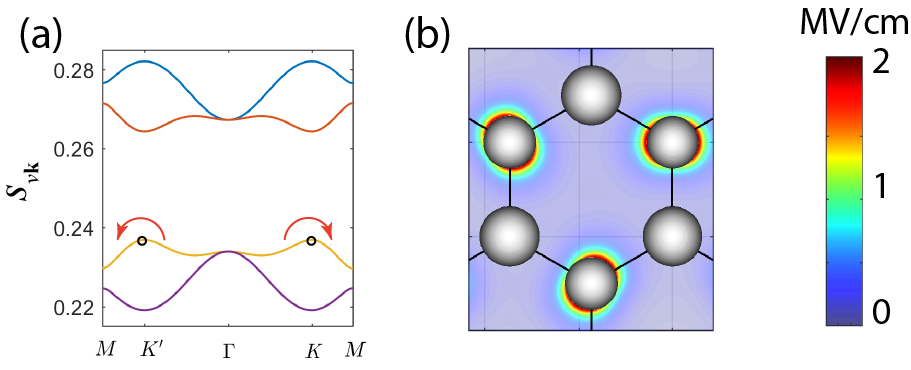}
\caption{(a) Band structure of the honeycomb array of metal nanoshells. Due to the broken inversion symmetry, i.e., sublattices A and B being inequivalent, there are band gaps at the $\mathbf  K$- and $\mathbf K^\prime $-points. The spasing SP modes at the $\mathbf K$- and $\mathbf K^\prime$-points are indicated by open circles; the arrows indicate the direction of the rotation of the local modal fields for the corresponding valleys.
(b) Profile of the electric field of an SP at the valence band in the $\mathbf K$-valley at a certain instance of time. Only the SPs at sublattice A are excited. The color bar to the right codes the modal field amplitude.}
\label{sp_field}
\end{figure}
\begin{figure}
\includegraphics[width=0.95\columnwidth]{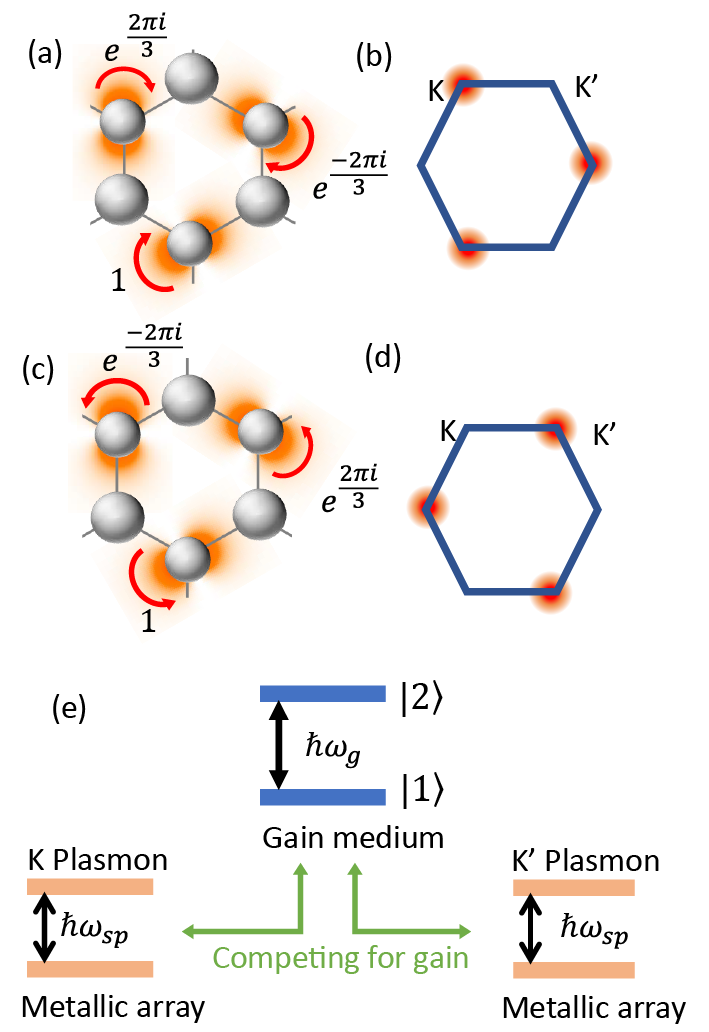}
\caption{(a)-(d) The valence band SPs in the $\mathbf K$- and $\mathbf K^\prime$-valleys. For both the valleys, only the SPs at the sublattice $A$ are excited. 
For the $\mathbf K$-valley, these are the SPs with $m=1$, while for the $\mathbf K^\prime $-valley, the SP with $m=-1$  are generated. Their phase shifts from site to site are $2\pi/3$ and 
$-2\pi/3$, respectively. The SP local fields rotate in the directions shown by the red arrows.(e) Schematic illustration of the system of gain medium and metal array. The SPs at the $\mathbf K$- and $\mathbf K^\prime $-points compete for the 
same gain.    }
\label{valley_SP}
\end{figure}

The solution of Eq.\ (\ref{tight_binding}) produces four bands whose dispersions are shown in Fig.\ \ref{sp_field}(a). Due to the broken inversion symmetry owing to the sublattices A and B being inequivalent, there are  band gaps opened at the $\mathbf K$- and $\mathbf K^\prime $-points.

The structure of SP 
excitations at these points is the following. At either the $\mathbf K$-point or the $\mathbf K^\prime$-point, the SP modes are excited in the real space only on a single sublattice,  $A$ -- see an illustration  in Fig.\ \ref{sp_field}(b). The SPs at the $\mathbf K$- and $\mathbf K^\prime$-points are mutually time-reversed: the magnetic quantum numbers are $m=\pm 1$  corresponding to the topological charges $Q_T=\pm1$, and the phase shifts between the nearest sites on the $A$ sublattice are $\pm2\pi/3$, respectively, as illustrated in Fig.\ \ref{valley_SP}(a)-(d). Below we assume that the 
frequency of the optical transitions in the gain medium is equal to the frequency of SPs in the valence band at the $\mathbf K$- and $\mathbf K^\prime$-points -- see a schematic in Fig.\ \ref{valley_SP}(e).

The SP's at the $\mathbf K$- and $\mathbf K^\prime$-points have equal frequencies, $\omega_{\nu\mathbf K}= \omega_{\nu\mathbf K^\prime }$, as protected by the $\mathcal T$ symmetry. We write down the Hamiltonian of the plasmonic system in the second quantization as   
\begin{align}
H_{SP}= \hbar\omega_{\nu\mathbf K} \left(\hat{a}_{\nu\mathbf{K}}^{\dagger}\hat{a}_{\nu\mathbf{K}}
+\hat{a}_{\nu\mathbf{K}^\prime}^\dagger\hat{a}_{\nu\mathbf{K^\prime}}\right),
\end{align}
where $\hat{a}_{\nu\mathbf{k}}^{\dagger}$ and $\hat{a}_{\nu\mathbf{k}}$ are the SP  creation and annihilation operators. Here the frequency $\omega_{\nu\mathbf{k}}$ is found from equation Re$[s(\omega_{\nu\mathbf{k}})]=s_{\nu\mathbf{k}}$, where $s(\omega)= \epsilon_d/(\epsilon_d-\epsilon_m(\omega))$ is the Bergman  spectral parameter, $\epsilon_d$ and $\epsilon_m(\omega)$ are the dielectric functions of the host material and the metal, respectively.  
The corresponding electric field is given by  \cite{Bergman_Stockman:2003_PRL_spaser} 
\begin{align}
&\mathbf{E}(\mathbf{r},t)=-\sum_{\mathbf{k}=\mathbf{K},\mathbf{K}^\prime}A_{\nu\mathbf{k}} \nabla \varphi_{\nu\mathbf{k}}(\mathbf{r})(\hat{a}_{\nu\mathbf{k}} +\hat{ a}_{\nu\mathbf{k}}^{\dagger} ),
\end{align}
 where $A_{\nu\mathbf{k}} =\sqrt{\left.4\pi\hbar s_{\nu\mathbf{k}}\right/{\epsilon_d s_{\nu\mathbf{k}}^\prime}} $, $s_{\nu\mathbf{k}}^\prime = \mathrm{Re}[{d s(\omega)}/{d\omega}|_{\omega=\omega_{\nu\mathbf{k}}}]$.

The gain medium is assumed to be achiral with a linear transition dipole moment $\mathbf{d}$; this dipole equally couples to both the $\mathbf K$ and $\mathbf K^\prime$ SP modes. The gain medium is described quantum mechanically using density matrix $\rho^{(p)}$ for a $p$th nanoshell. Within the rotating wave approximation (RWA),  the non-diagonal elements of the density matrix can be written as $(\rho^{(p)})_{12}=\bar{\rho}^{(p)}\exp(i\omega_{\nu\mathbf K}t)$, while the diagonal elements determine the population inversion, $n^{(p)}=\rho^{(p)}_{22} - \rho^{(p)}_{11}$.
The interaction between the gain medium and the SP system is determined by Hamiltonian 
\begin{equation}
H_{int} = -\sum_p \mathbf{E}(\mathbf{r}^{(p)},t) \mathbf{d}^{(p)}.
\end{equation}

Following Ref.~\onlinecite{Stockman_JOPT_2010_Spaser_Nanoamplifier}, we treat the plasmonic field quasiclassically by replacing the creation and annihilation operators of the SP's by the respective complex $c$-number amplitudes $a_{\nu\mathbf K}^\ast$, $a_{\nu\mathbf K}$, and $a_{\nu\mathbf K^\prime}^\ast$, $a_{\nu\mathbf K^\prime}$. The corresponding SP population numbers per the composite unit cell are $N_{\mathbf K}=\left|a_{\nu\mathbf K}\right|^2$ and $N_{\mathbf K^\prime}=\left|a_{\nu\mathbf K^\prime}\right|^2$.

The coupled system of equations \cite{Stockman_JOPT_2010_Spaser_Nanoamplifier}, which describes both the SPs and the gain medium, takes the following form  
\begin{eqnarray}
\dot{a}_{\nu \mathbf K} & = & -\gamma_\mathrm{sp}{a}_{\nu  \mathbf K }+ i N_\mathrm{g}\sum_{p }\bar{\rho}^{(p)*}\tilde{\Omega}_{\nu \mathbf K}^{(p)*}, \label{eq:hsp_2s_1}\\
 \dot{a}_{\nu \mathbf K^\prime} & = & -\gamma_{sp}{a}_{\nu  \mathbf K^\prime} + i N_\mathrm{g}\sum_{p}\bar{\rho}^{(p)*}\tilde{\Omega}_{\nu \mathbf K^\prime}^{(p)*}, \label{eq:hsp_2s_2}\\
\dot{n}^{(p)} & = &-4\mathrm{Im} \left[ \bar{\rho}^{(p)} \sum_{\mathbf{k}=\mathbf K, \mathbf K^\prime}\tilde{\Omega}_{\nu \mathbf{k}}^{(p)}a_{\nu \mathbf{k}}^{(p)}\right]  + \nonumber \\
& & g (1-n^{(p)})- \gamma_{2}(1+n^{(p)}), \label{eq:hsp_2s_3}\\
\dot{\bar{\rho}}^{(p)} & = & -\Gamma_{12} \bar{\rho}^{(p)}+i n^{(p)}\sum_{\mathbf{k}=\mathbf K,\mathbf K^\prime } \tilde{\Omega}_{\nu \mathbf{k}}^{(p)*}a_{\nu \mathbf{k}}^{(p)*} ,\label{eq:hsp_2s_4}
\end{eqnarray}
 where $N_g$ is the number of electrons of the gain medium inside each shell, $\gamma_{sp}$ is the SP relaxation rate,  $\gamma_{2}$ is the non-radiative decay rate of the level $|2\rangle$ of the chromophore, $\Gamma_{12}$ is polarization relaxation rate for the $|2\rangle\to|1\rangle$ transition of the chromophores, $g$ is the excitation (pumping) rate per a chromophore, and $\tilde{\Omega}_{\nu \mathbf{k}}^{(p)}= \frac{1}{\hbar}A_{\nu\mathbf{k}} \nabla \varphi_{\nu\mathbf{k}}(\mathbf{r}^{(p)})\mathbf d^{(p)}$ is the  Rabi frequency. In computations, we set: $\epsilon_d=2$, $d=10$ debye, $ N_\mathrm g=514$, $\gamma_\mathrm{sp}=4.1\times10^{13}$ $\mathrm{s^{-1}}$,  $\Gamma_{12}=2.1\times10^{14}$ $\mathrm{s^{-1}}$, and $\gamma_2=4\times10^{12}$ $\mathrm{s^{-1}}$. We used dielectric data \cite{Johnson:1972_Silver} for silver as the nanoshells' metal.
 
The Rabi frequencies, $\tilde{\Omega}_{\nu \mathbf K}^{(p)}$ and $\tilde{\Omega}_{\nu \mathbf K^\prime }^{(p)}$, are periodic with respect to the position vector, $\mathbf{r}^{(p)}$, with the periods that are determined by the crystal momenta $\mathbf K$ and $\mathbf K^\prime$, respectively. Any solution of Eqs. (\ref{eq:hsp_2s_1})-(\ref{eq:hsp_2s_4}), which contains both the $\mathbf K$ and $\mathbf K^\prime$ components, will be periodic on a superlattice with an eighteen-nanoshell unit cell (``supercell'') shown in Fig. \ref{fig:honeycomb_array}. Then the summation over $p$ in Eqs.\ (\ref{eq:hsp_2s_1}) and (\ref{eq:hsp_2s_2}) is extended over the corresponding eighteen points. 

\begin{figure}
\includegraphics[width=.95\columnwidth]{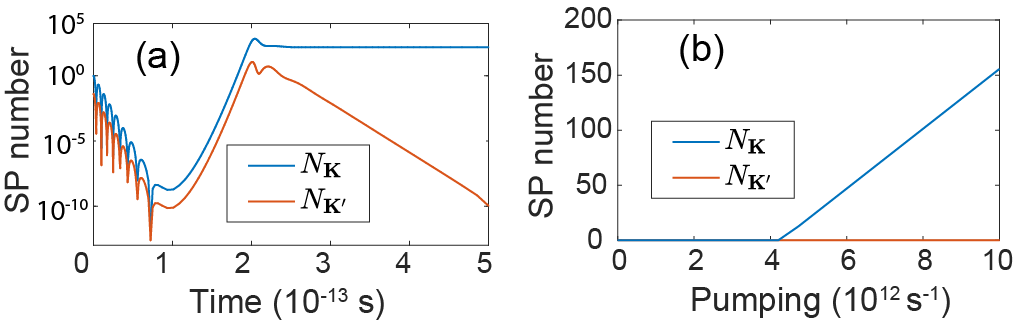}
\caption{Dynamics of SP population per unit supercell. (a) Temporal dynamics of the SP population. Initially the SPs in both the valleys are present, which is case (iii) -- see the text. The pumping rate is $g=10^{13}$ $\mathrm{s^{-1}}$. (b) The stationary number of SPs in both valleys as a function of pumping rate.
}
\label{sp_number_dynamics}
\end{figure}

\begin{figure}
\includegraphics[width=0.95\columnwidth]{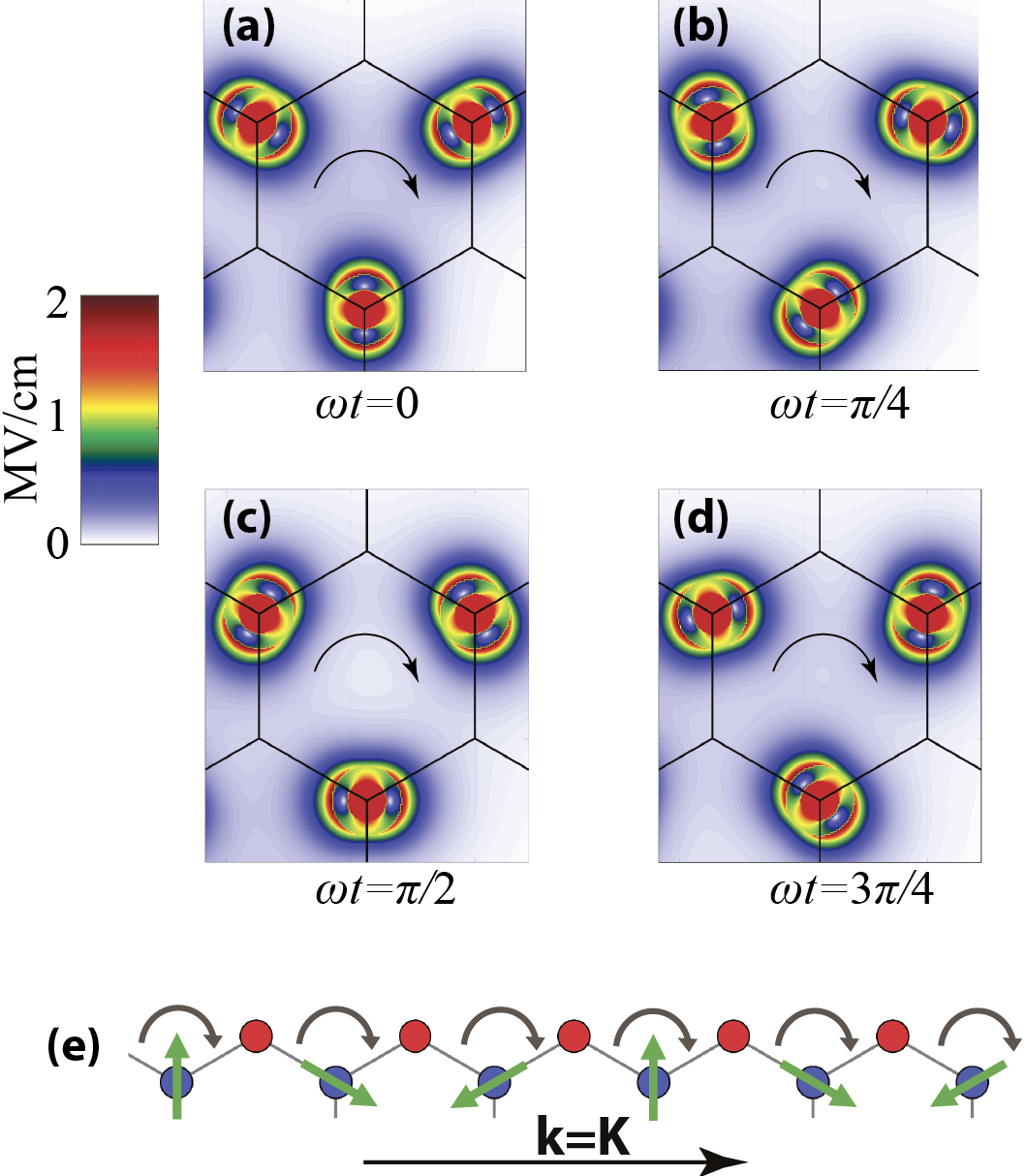}
\caption{Dynamics of $\mathbf K$-valley SP eigenmode. (a)-(d) The distributions of the local field modulus are plotted for the unit cell in the real space. The phases of the spaser oscillation are indicated for each panel. The color-coded scale of the local fields is given at the left-hand side for one SP quantum per the unit supercell. The local fields rotate in time  in the direction of the black arrow, i.e., clockwise. (e) Edge fields for $\mathbf K$-valley mode. The instantaneous dipole vectors are indicated by the green arrows; the direction of the field rotation is depicted by the curved black arrows;  crystal momentum $\mathbf k$ of the mode is shown by the straight black arrow. Each such a rotating dipole generates a current that is normal to this dipole.
}
\label{Spaser_Field_Dynamics}
\end{figure}

As initial conditions for Eqs.\ (\ref{eq:hsp_2s_1})-(\ref{eq:hsp_2s_4}), we set $n^{(p)}(t)\vert_{t=0}=-1$, which implies that initially all the chromophores of the gain medium are  in the ground state. Similarly, we assume that initially there is no polarization, $\bar\rho(t)\vert_{t=0}=0$. With respect to the initial SP amplitudes, $a_{\nu\mathbf K}$ and $a_{\nu\mathbf K^\prime}$, we consider three cases: (i) The initial SP amplitude is small and located only at the $\mathbf K$-point where we set $a_{\nu\mathbf K}=0.1$ and $a_{\nu\mathbf K^\prime}=0$; (ii) The same as the previous case but with the SP amplitude located at the $\mathbf K^\prime$-point, $a_{\nu\mathbf K}=0$ and $a_{\nu\mathbf K^\prime}=0.1$; (iii) The SP's are initially present at both the $\mathbf K$- and $\mathbf K^\prime$-points but with different amplitudes, $a_{\nu\mathbf K}=1$ and $a_{\nu\mathbf K^\prime}=0.2$.

We start with case (iii) where initially both the  $\mathbf K$- and $\mathbf K^\prime$-valleys are weakly populated, with a higher population at the $\mathbf K$-point, which emulates a random initial condition created by quantum fluctuations.  The pumping begins at $t=0$ with a rate of $g= 10^{13}$ $\mathrm{s^{-1}}$. The dynamics of the SP populations in both the valleys as a function of time $t$ after the beginning of the pumping is displayed in Fig.\ \ref{sp_number_dynamics}(a). As we see, after the initial period of decay and relaxation oscillations, the population of the dominant $\mathbf K$-valley starts growing exponentially and then levels off becoming stationary with the SP population $N_{\mathbf K}\approx 150$. At the same time, the minor SP population, $N_{\mathbf K^\prime}$,  decays exponentially for $t\gtrsim 200$ fs. This occurs due to a strong competition of the $\mathbf K$- and $\mathbf K^\prime$-modes for the common gain. At the same time, the decay to zero, $N_{\mathbf K^\prime}\to0$, implies that there is no cross-mode talk, which is due to the topological protection: these two SP modes carry conserved topological charges $Q_T=\pm1$. Due to the $\mathcal T$-symmetry, the valleys are symmetric: starting with the $\mathbf K^\prime$-valley with a dominant SP population, the final population will only be in that valley. Thus the topological spaser is a topologically-protected symmetric bistable device.

The kinetics of the topological spaser, i.e., the dependence of the stationary SP population numbers,  $N_{\mathbf K}$ and  $N_{\mathbf K^\prime}$ for $t\to\infty$, as functions of the pumping rate $g$, is illustrated in Fig.\ \ref{sp_number_dynamics}(b) for case (i) when only the $\mathbf K$-valley has a non-zero initial population: $N_\mathbf K=0.01$ and $N_{\mathbf K^\prime}$=0. There is a pronounced threshold after which the SP population in the $\mathbf K$-valley grows linearly with the pumping rate. The population in the $\mathbf K^\prime$-valley remains zero due to the topological protection. Similarly, in case (ii), the SP population will be amplified and remain only in the $\mathbf K^\prime$-valley.

The dynamics of the real-space distributions of the local fields in the topological spaser is shown in Fig.\ \ref{Spaser_Field_Dynamics} (a)-(e) for the case of the $\mathbf K$-valley generation (see also  Appendix for the $\mathbf K^\prime$ mode). The fields are normalized to one SP in the unit supercell. This distribution contains three dipolar fields localized on the A-sublattice nanoshells, which rotate clockwise in the optical cycle. This field distribution is generally chiral and compliant with the $E^{\prime}$ representation of the $D_{3h}$ group. At a given moment of time, this field distribution can also be described as a chiral lattice wave propagating in the clockwise direction along the unit cell boundary. The 2d $E^{\prime}$ representation also describes a mode with the opposite (counter-clockwise) chirality, which can be obtained by an application of either $\mathcal T$ or $\sigma_v^\prime$ symmetry operations.

The present topologically-charged $E^{\prime}$ spasing mode is indeed dark: obviously, it does not possess a net dipole moment. However, there is a non-zero chiral current propagating unidirectionally clockwise (for $Q_T=1$) or counterclockwise (for $Q_T=-1$) within the A-sublattice unit cell. Such currents for the neighboring cells cancel out each other. However, there will be an uncompensated chiral current at the edge of the 2d plasmonic lattice, which is illustrated in  Fig.\ \ref{Spaser_Field_Dynamics} (e). The corresponding local fields resemble a field of a plasmon polariton propagating clockwise along the edge despite the absence of the edge states, similar to the edge plasmon polaritons in Ref.\ \cite{Song_et_al_PNAS_2016_Chiral_plasmons_without_magnetic_field}.

There is the second $E^{\prime}$ mode of the same frequency, which is related to the above-discussed right-rotating one by the application of either $\mathcal T$ or $\sigma_v^\prime$ symmetry operations. That mode is generated in the $\mathbf K^\prime$-valley; it is rotating counter-clockwise within the unit cell and propagating counter-clockwise along the edge. The choice of which one of these modes will be generated is random and is a spontaneous violation of the $\mathcal T$ and $\sigma_v^\prime$ symmetries.

In conclusion, we propose a topological spaser constituted by a honeycomb lattice of spherical metal nanoshells containing a gain medium. The two sublattices, A and B, are built from two different types of nanoshells. Such a spaser generates one of two chiral plasmonic modes characterized by topological charge $Q_T=\pm1$ whose local fields rotate clockwise or counter-clockwise, respectively.  Which of these two modes is generated is determined by a spontaneous violation of symmetry as defined by the initial conditions. The chirality of the spasing mode is stable and topologically protected. The macroscopic fields of the spasing mode are localized along the edge of the lattice despite the absence of the edge states. These edge fields propagate clockwise (for $Q_T=1$) or counter-clockwise ($Q_T=-1$). Due to a very high lattice momentum $\mathbf k=\mathbf K$ or $\mathbf k=\mathbf K^\prime$, they are dark in contrast to the original lasing spaser of Ref.\ \cite{Zheludev_et_al_Nat_Phot_2008_Lasing_Spaser}. Nevertheless, they can be outcoupled by using a corresponding coupler, e.g., grating. Application-wise, such a topological spaser is a symmetric bistable that can be used for ultrafast optical storage and processing of information.

\begin{acknowledgments}
Major funding was provided by Grant No. DE-FG02-11ER46789 from the Materials Sciences and Engineering Division of the Office of the Basic Energy Sciences, Office of Science, U.S. Department of Energy. Numerical simulations have been performed using support by
Grant No. DE-FG02-01ER15213 from the Chemical Sciences, Biosciences and Geosciences Division, Office of Basic Energy Sciences, Office of Science, US Department of Energy. The work of V.A. was supported by NSF EFRI NewLAW Grant EFMA-17 41691. Support for J.W. came from a MURI Grant No. N00014-17-1-2588 from the Office of Naval Research (ONR).
\end{acknowledgments}
\appendix
\section{Dipole Modes of metallic spherical shell}
The surface plasmon eigenmodes of the $i$th nanoshell  are described by the equation

\begin{align}
\nabla\left[\theta^{(i)}(\mathbf{r})\nabla \varphi_m^{(i)} \right]= s_m^{(i)} \nabla^2 \varphi_m^{(i)}. 
\end{align}
Function  $\theta^{(i)}(\mathbf{r})$ of the nanoshell is equal to 1 inside the metal of the $i$th nanoshell and 0 elsewhere. 
The $m=\pm1$ dipole eigenmode is given by
\begin{align}
\varphi = \sin\theta e^{im\phi}
\begin{cases}
c_1 \frac{r}{R_1},& 0<r<R_1,\\
c_2 \frac{r}{R_1},+c_3\frac{R_1^2}{r^2},& R_1<r<R_2,\\
c_4 \frac{R_1^2}{r^2},& R_2<r.\\
\end{cases}\label{eq:c_dipole}
\end{align}
One finds  the eigenvalue $s^{(i)}_m$ for these eigenmodes as
\begin{align}
s_m^{(i)}  &= \frac{1\pm\frac{1}{3}\sqrt{1+8\eta^3}}{2},
\label{s_m}
\end{align}
where
\begin{equation}
\eta = \frac{R_1}{R_2}~.
\end{equation}
We impose a normalization condition
\begin{align}
\int_{\mathrm{All~Space}} |\nabla \varphi_m^{(i)} |^2 dv =1~.
\end{align}
There are two solutions corresponding to the $\pm$ signs in Eq.\ (\ref{s_m}). We choose one corresponding to the $-$ sign, which describes a lower frequency mode. Substituting it, we find:
\begin{widetext}
\begin{align}
c_1 &=\left[ \frac{3}{16\pi R_1}\frac{(1-s_m^{(i)})(s_m^{(i)}-\frac{1}{3})}{(s_m^{(i)}-\frac{1}{2})}
\right]^{1/2} =\left[ \frac{-1+4\eta^3+\sqrt{1+8\eta^3}}{16 \pi R_1\sqrt{1+8\eta^3}}
\right]^{1/2},\\
c_2 &=\left(\frac{s_m^{(i)}-\frac{2}{3}}{s_m^{(i)}-1}\right)c_1,\\
c_3 &=\frac{1}{3}\left(\frac{-1}{s_m^{(i)}-1}\right)c_1,\\
c_4 &=\frac{1}{3}\left(\frac{-1}{s_m^{(i)}-\frac{1}{3}}\right)c_1.
\end{align}
\end{widetext}
The electric field of the SP mode inside the gain core, i.e., for $r\le R_1$, is 
\begin{align}
\mathbf{E} &= -A\frac{c_{1} e^{ im\phi}}{R_1}(\mathbf e_{r}\sin\theta+\mathbf e_{\theta}\cos\theta+ i\mathbf e_{\phi})\notag\\
&= -A\frac{c_{1} }{R_1}(\mathbf e_{x}+ im\mathbf e_{y}),
\end{align}
where $\mathbf e_r, \mathbf e_\theta, \mathbf e_\phi$ and $\mathbf e_x, \mathbf e_y$ are the corresponding spherical and Cartesian unit vectors, and
\begin{align}
A & =\sqrt{ \frac{4\pi\hbar s_m^{(i)}}{\epsilon_h s_m^{(i)\prime} }}.
\end{align}

\section{SPP of Plasmonic Array}
Consider the problem of two media with permittivities $\epsilon_d$ and $\epsilon_m$.
The eigenmodes of SP's  in the quasistatic limit can be obtained from 
\begin{align}
\nabla\left[\theta(\mathbf{r})\nabla \varphi\right] = s \nabla^2 \varphi, 
\end{align}
with $s={\epsilon_d}/({\epsilon_d-\epsilon_m})$ and $\theta(\mathbf{r})$ is  1 inside metal and zero elsewhere.
The geometric properties of the eigenmodes are characterized by a generalized eigenvalue problem,
\begin{align}
\nabla\left[\theta(\mathbf{r})\nabla \varphi_n\right] = s_n \nabla^2 \varphi_n, \label{eq:field_eq}
\end{align}
where $s_n$ and $\varphi_n$ are the eigenvalues and eigenfunctions, respectively.
The conditions on the boundary $S$ are
\begin{align}
\varphi_n \cdot\mathbf{n}\nabla\varphi_n =0,
\end{align}
where $\mathbf{n}$ is the normal vector to $S$. 
Multiplying Eq.~\eqref{eq:field_eq} by $\varphi_n$ and integrating by parts, we obtain
\begin{align}
s_n &= \langle n|\theta(\mathbf{r})| n\rangle,\\
|n\rangle &= \nabla \varphi_n,\\
\langle m|n\rangle &= \int d^3r \left(\nabla \varphi_m^*\right)\left(\nabla \varphi_n\right),   
\end{align}
The eigenvectors are orthogonal and normalized as,
\begin{align}
\langle m|n\rangle &= \delta_{mn}.   
\end{align}
Consider that there are $N$ metal particles (same metal).  
\begin{align}
&\nabla\left[\theta(\mathbf{r})\nabla \varphi_n\right] = s_n \nabla^2 \varphi_n,\label{eq:n_particle}\\
&\theta\mathbf{r})=\sum_{i=1}^{N}\theta^{(i)}(\mathbf{r}).
\end{align} 
We can solve Eq.~\eqref{eq:n_particle} in terms of the local basis $\varphi_{m}^{(i)}$, defined as 
\begin{align}
&\nabla\left[\theta^{(i)}(\mathbf{r})\nabla \varphi_{m}^{(i)}\right] = s_{m}^{(i)} \nabla^2 \varphi_{m}^{(i)},\label{eq:local_basis}
\end{align} 
where $\varphi_{m}^{(i)}$ are the eigenmodes of the $i$th metal particle. 
We expand the total field $\varphi_n$ in the local basis
\begin{align}
\varphi_n = \sum_{i ,m}C_{nm}^{(i)}\varphi_{m}^{(i)}\label{eq:l_expansion},
\end{align}
where $C_{nm}^{(i)}$ are the coefficients to be determined. 
Substituting Eq.~\eqref{eq:l_expansion} in  Eq.~\eqref{eq:n_particle} and using Eq.~\eqref{eq:local_basis}
we obtained
\begin{widetext}
\begin{align}
\sum_{j}\sum_{i\neq j,m}C_{nm}^{(i)}\nabla\left[\theta_j(\mathbf{r})\nabla \varphi_{m}^{(i)}\right]
+ 
\sum_{i,m}s_{m}^{(i)}C_{nm}^{(i)}\nabla^2\varphi_{m}^{(i)}= s_n \sum_{i,m}C_{nm}^{(i)}\nabla^2\varphi_{m}^{(i)},
\label{eq:l_expression_0}
\end{align}
\end{widetext}
where the first term describes the inter-particle interaction.
To deal with the first term, we expand the field $\nabla\varphi_{m}^{(i)}$  
 over the other local eigenstates $\nabla\varphi_{m'}^{(j)}$ ($i\neq j$). The field $\nabla\varphi_{m}^{(i)}$  
inside the $j$th particle can be expanded over the local basis $\nabla\varphi_{m'}^{(j)}$, since  
the local basis $\nabla\varphi_{m'}^{(j)}$ forms a complete set of field inside the $j$th metal.
\begin{align}
\nabla\varphi_{m}^{(i)}=\sum_{ m^{\prime}}\beta_{ijmm'}\nabla\varphi_{m'}^{(j)}
\label{eq:l_expression_1}.
\end{align}
The interaction can be written as
\begin{align}
\sum_{j,m'}\sum_{i\neq j,m}C_{nm}^{(i)}\beta_{ijmm'}\nabla\left[\theta_j(\mathbf{r})\nabla \varphi_{m'}^{(j)}\right]
\end{align}
or equivalently
\begin{align}
\sum_{j,m'}\sum_{i\neq j,m}C_{nm}^{(i)}\beta_{ijmm'}s_{m'}^{(j)}\nabla^2\varphi_{m'}^{(j)}.
\end{align}
The eigenmode equation becomes
\begin{widetext}
\begin{align}
\sum_{j,m'}\sum_{i\neq j,m}C_{nm}^{(i)}\beta_{ijmm'}s_{m'}^{(j)}\nabla^2\varphi_{m'}^{(j)}
+ 
\sum_{i,m}s_{m}^{(i)}C_{nm}^{(i)}\nabla^2\varphi_{m}^{(i)}= s_n \sum_{i,m}C_{nm}^{(i)}\nabla^2\varphi_{m}^{(i)}.
\end{align}
\end{widetext}
Multiplying $\phi_{m}^{(j)}$ on the both sides,  integrating by parts and utilizing the othogonality of the eigenmodes, we obtain
\begin{align}
\sum_{i\neq k,m'}C_{nm'}^{(i)}\beta_{ikm'm}s_{m}^{(k)}
=  (s_n-s_{m}^{(k)})C_{nm}^{(k)}
\end{align}
More compactly,
\begin{align}
\sum_{i,m'}\beta_{ikm'm}s_{m}^{(k)}C_{nm'}^{(i)}
=  s_nC_{nm}^{(k)}. \label{eq:SPP_array_main}
\end{align}
Equation \eqref{eq:SPP_array_main} defines an eigenproblem, which yields the egienvalue $s_n$ and the coefficients $C_{nm}^{(k)}$.  
The Hermiticity of $\beta_{ijmm'}$, i.e.,
\begin{align}
\beta_{ijmm'}=\beta_{jim'm}
\end{align}
 is protected by the Green's reciprocity.
The interaction coefficeints $\beta_{ijmm'}$ are given by an integral
\begin{widetext}
\begin{align}
\beta_{ijmm'}& =\frac{1}{s_{m'}^{(j)}}\int \theta^{(j)}(\mathbf{r})\left( \nabla \varphi_{m' }^{(j)*}\right)\left(\nabla\varphi_{m}^{(i)}\right) d^3r=\frac{1}{s_{m}^{(i)}}\int \theta^{(i)}(\mathbf{r}) \left(\nabla \varphi_{m' }^{(j)*}\right)\left(\nabla\varphi_{m}^{(i)}\right) d^3r\\
                    & =\int_{\mathrm{All~Space}} \left(\nabla \varphi_{m' }^{(j)*}\right)\left(\nabla\varphi_{m}^{(i)}\right) d^3r.\label{eq:overplap}
\end{align}
\end{widetext}
Because matrix $\beta_{ikm'm}s_{m}^{(k)}$ in Eq.~\eqref{eq:SPP_array_main} is not Hermitian,  the eigenvectors are not orthogonal, and  the normalization condition is not $\sum_{i,m} |C_{nm}^{(i)}|^2=1 $.
In actuality, the normalization condition is given by
\begin{align}
\sum_{i,k,m',m}C_{nm}^{(k)*}\beta_{ikm'm}C_{nm'}^{(i)}
= 1.\label{eq:normalization}
\end{align}
We can convert Eq.~\eqref{eq:SPP_array_main} into a Hermitian eigenproblem by
a substitution, $\tilde{C}_{nm}^{(i)}=\sqrt{s_m^{(i)}}C_{nm}^{(i)}$.
With this substitution, we rewrite Eq.~\eqref{eq:SPP_array_main} as
\begin{align}
\sum_{i,m'}H_{ikm'm}\tilde{C}_{nm'}^{(i)}
=  s_n\tilde{C}_{nm}^{(k)}~, 
\label{eq:SPP_array_her}\\
H_{ikm'm}=\sqrt{s_m^{(i)}s_{m'}^{(j)}}\beta_{ikm'm}~.
\end{align}
Because $H_{ikm'm}$ is Hermitian, the eigenvectors are orthogonal and the normalization is simply
\begin{align}
\sum_{i,m}|\tilde{C}_{nm}^{(i)}|^2
= 1.\label{eq:normalization_her}
\end{align}


\section{Characters of Topological SP Eigenmodes as  Irreducible Representations of the $D_{3h}$ Point Symmetry Group}

The spasing eigenmodes are surface plasmons (SPs) that should be classified corresponding to irreducible representations of the point symmetry group of the lattice unit cell, which is $D_{3h}$  \cite{Landau_Lifshitz_Quantum_Mechanics:1965}. This group has six such representations: four of them are singlet or one-dimensional (1d), which are $A_1^\prime$ (scalar or fully-symmetric), $A_2^\prime$ (a pseudovector in the $z$-direction),  $A_1^{\prime\prime}$ (pseudoscalar), and $A_2^{\prime\prime}$ (a polar vector in the $z$-direction), and two are doublet or 2d-representations: $E^\prime$ (a polar vector in the $xy$-plane) and $E^{\prime\prime}$ (a pseudovector in the $xy$-plane). We seek for a solution where the local electric field is in the plane of the lattice (the $xy$-plane) and rotates in time for a given lattice site and in space from a site to a site. A basis for such a field can be constituted, e.g., by two spherical harmonics, $Y_{1\pm 1}$ corresponding to left- and right-handed rotations. 

\begin{figure}
\includegraphics[width=0.5\columnwidth]{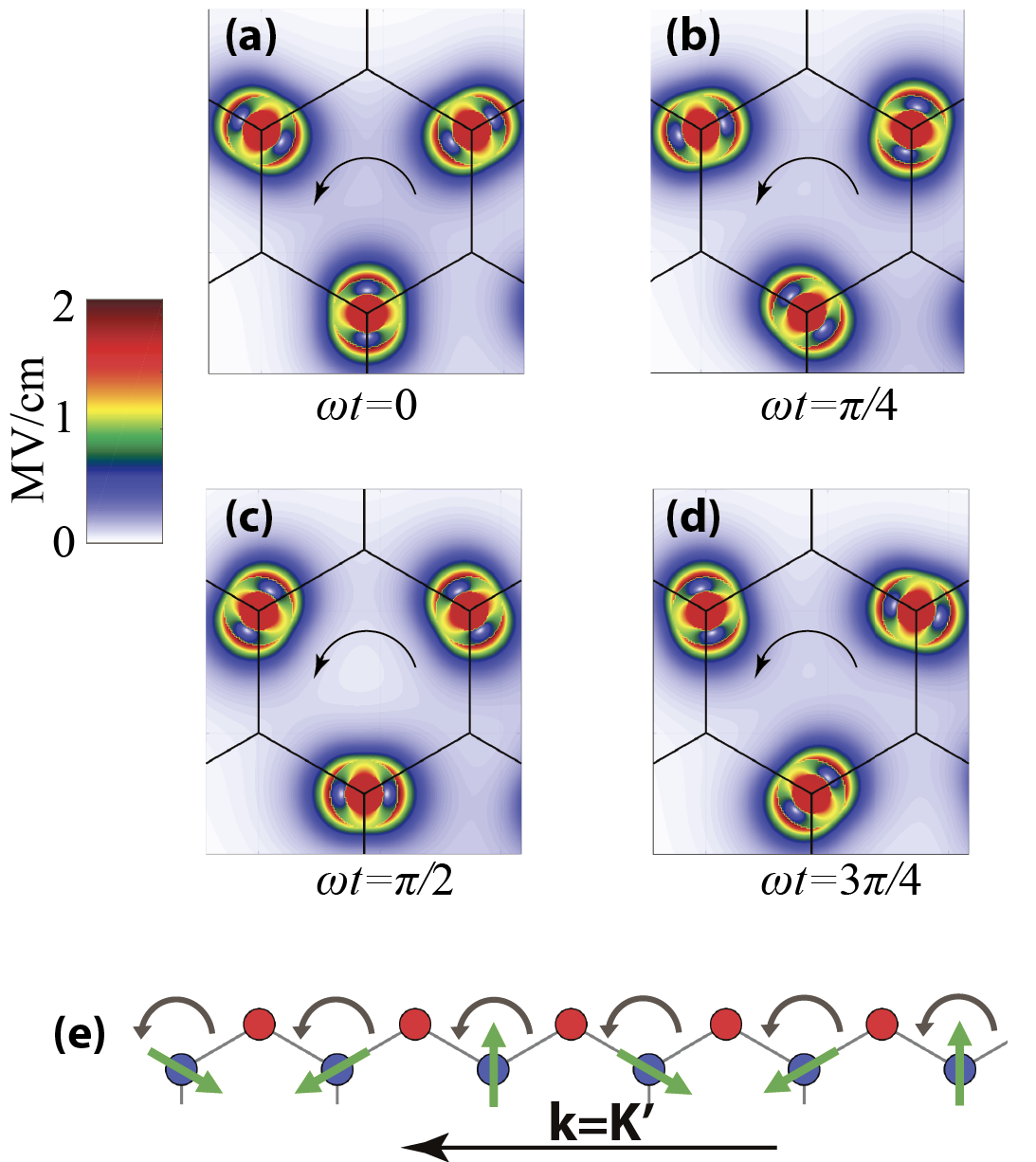}
\caption{Dynamics of the $\mathbf K^\prime$-valley SP eigenmode. Same as Fig.\ \ref{Spaser_Field_Dynamics} but the dynamics is $\mathcal T$-reversed: fields rotate left and propagate left.
\label{Spaser_Field_Dynamics_L}
}
\end{figure}

An example of the local fields of the two eigenmodes, which are mutually time-reversed and should be described by the same doublet representation, is provided by our solution of the spaser equations -- see the main text of the article. Here we show such two modes: the $\mathbf K$-mode rotating right [Fig.\ \ref{Spaser_Field_Dynamics}] and the $\mathcal T$-reversed $\mathbf K^\prime$-mode rotating left [Fig.\ \ref{Spaser_Field_Dynamics_L}].

To unambiguously identify the corresponding representation, we note that the underlying material system is a sublattice A whose unit cell has the third-oder vertical ($z$) axis ($3C_3$ symmetry elements), three second-order axes (in the $xy$ plane) ($3U_2$ symmetries), also $3S_3$ elements (mirror-rotational symmetries), and $3\sigma_v^\prime$ elements (reflections in the vertical planes). 
Spherical harmonics $Y_{1\pm1}$ are spherical representations of a polar vector in the $xy$ plane: 
\begin{equation}
Y_{1\pm1}=\mp\frac{1}{2}\sqrt{\frac{3}{2\pi}}(x\pm iy)~.
\label{Y1pm1}
\end{equation}

\begin{figure}
\includegraphics[width=0.5\columnwidth]{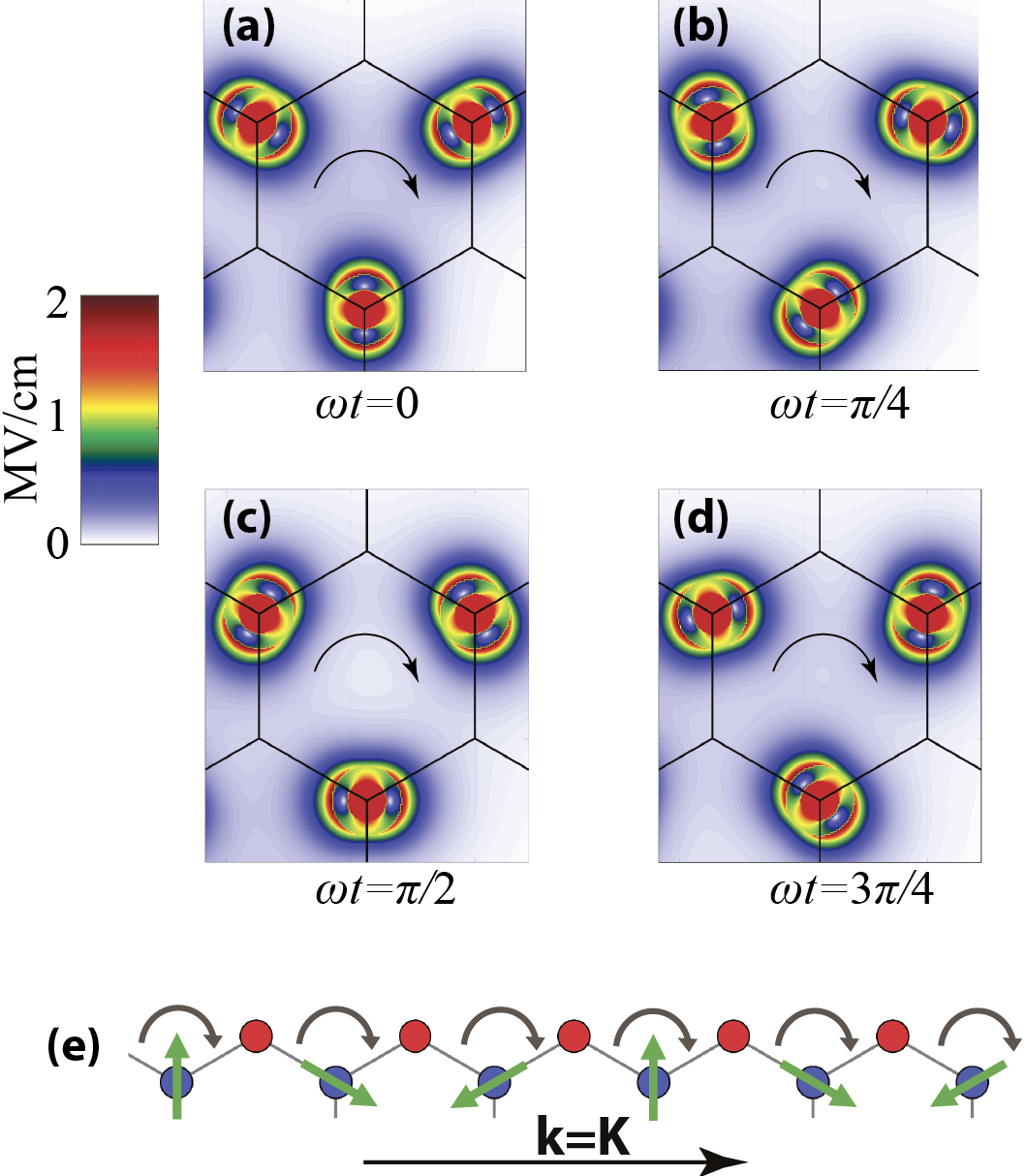}
\caption{Dynamics of $\mathbf K$-valley SP eigenmode. (a)-(d) The distributions of the local field modulus are plotted for the unit cell in the real space. The phases of the spaser oscillation are indicated for each panel. The color-coded scale of the local fields is given at the left=hand side for one SP quantum per the unit supercell. The local fields rotate in time  in the direction of the black arrow, i.e., clockwise. (e) Edge fields for $\mathbf K$-valley mode. The instantaneous dipole vectors are indicated by the green arrows; the direction of the field rotation is depicted by the curved black arrows;  crystal momentum $\mathbf k$ of the mode is shown by the straight black arrow. Each such a rotating dipole generates a current that is normal to this dipole.
\label{Spaser_Field_Dynamics}
}
\end{figure}
\begin{table}[]
\begin{tabular}{@{}ccccccc@{}}
\hline
                              & $E$                   & $\sigma_h$ & $3C_3$  & $3S_3$  & $3U_2$  &  $3\sigma_v^{\prime}$  \\ \hline
$E^{\prime}$ &    2                   &  2              & -1         & -1           &  0          &    0   \\ \hline
\end{tabular}
\caption{Characters of the $E^{\prime}$ representation of the $D_{3h}$ point symmetry group, adapted from Ref. \onlinecite{Landau_Lifshitz_Quantum_Mechanics:1965}.
The elements of $D_{3h}$ are the identity $E$, the horizontal-plane reflection $\sigma_h$, the three rotations $C_3$, the three rotary-reflections $S_3$, 
the three rotations about the horizontal axes $U_2$, and the three vertical-plane reflections $\sigma_v^{\prime}$.
}\label{tb:ch}
\end{table}

The character of the trivial $E$ operation is always the representation dimensionality, i.e., $\chi(E)=2$. To find $\chi(\sigma_h)$ for our eigenvectors we note that the $\sigma_h$ operation is reflection in the $xy$ mirror plane (i.e., a transformation $x\to x$, $y\to y$, and $z\to -z$), which leaves both of the basis functions of Eq.\ (\ref{Y1pm1}) unchanged. Correspondingly, $\chi(\sigma_h)=2$. Operation $C_3$ is a rotation by a $2\pi/3$ angle whose character, obviously, is $\chi(C_3)=2\cos\left(2\pi/3\right)=-1$. An operation $S_3$ is a product of $\sigma_h$, which leaves the basis unchanged, and $C_3$; thus its character is $\chi(S_3)=\chi(C_3)=-1$. A $U_2$ rotation, say about the $y$ axis of symmetry, causes $y\to y, x\to -x$; this results in $Y_{11}\leftrightarrow Y_{1-1}$. Thus the basis functions are exchanged, and the corresponding transformation matrix is purely off-diagonal; correspondingly, $\chi(U_2)=0$. An operation $\sigma_v^\prime$ acts for our purposes exactly like $U_2$; thus $\chi(\sigma_v^\prime)=0$. These characters  that we have obtained are exactly those of the $E^\prime$ representation of the $D_{3h}$ group summarized in Table \ref{tb:ch}. This unambiguously identifies that our spasing eigenmodes transform accordingly to the $E^\prime$ doublet representation of the $D_{3h}$ point symmetry group. Note that that the two components of this doublet describe field distribution rotating in time and space in the opposite directions. They transform one into another by either of the following transformations: $\mathcal T$, $\sigma_v^\prime$, or $U_2$.

%
%

\end{document}